 \documentclass[nohyper,12pt,letterpaper]{JHEP3}

\usepackage{amsfonts,amsmath,amsopn,amssymb,amsthm,bbm,latexsym,mathrsfs,verbatim}
\usepackage[dvipdfm,dvips]{graphicx}

 \title{Undetected Higgs decays and neutrino masses in gauge mediated, lepton number
 violating models}
 \author{T.\,Banks\\
 Department of Physics and SCIPP\\
 University of California, Santa Cruz, CA 95064\\
 E-mail: \email{banks@scipp.ucsc.edu}\\
 {\it and}\\
 Department of Physics and NHETC, Rutgers University\\
 Piscataway, NJ 08540 }

 \author{L.M.\,Carpenter\\
 Department of Physics and SCIPP\\
 University of California, Santa Cruz, CA 95064\\
 E-mail: \email{lmc@scipp.ucsc.edu}
  }

 \author{J.-F.\,Fortin\\
 Department of Physics and NHETC, Rutgers University\\
 Piscataway, NJ 08540 \\
 E-mail: \email{jffor27@physics.rutgers.edu}
  }

 \abstract{We discuss SUSY models in which renormalizable lepton number violating
 couplings hide the decay of the Higgs through $h\rightarrow\chi_1^0\chi_1^0$ followed
 by $\chi_1^0\rightarrow\tau jj$ or $\chi_1^0\rightarrow\nu_\tau jj$ and also explain neutrino
 masses.  This mechanism can be made compatible with gauge mediated
 SUSY breaking.
 }
 \received{} \accepted{}
 \preprint{hep-th{}\\\\ \\}
 
\begin{document}

\section{The little hierarchy problem and its solutions}

The minimal supersymmetric extension of the standard model (MSSM)
predicts a light Higgs boson.  While theory predicts a tree level
Higgs mass which is at most the mass of the Z boson, the current
experimental lower bound from LEP \cite{LEPbound} is $114.4$ GeV.
Evading the experimental lower bound requires significant one loop
corrections which can be achieved only by fine tuning of parameters
\cite{ef}.   This {\it little hierarchy problem}, while nowhere near
as severe as the original gauge hierarchy problem, has excited a lot
of theoretical interest.  A variety of solutions has been proposed
\cite{solutions}.  Some of them introduce new degrees of freedom to
enhance the contributions to the Higgs mass, while others allow for
non-standard decays of the Higgs, which would have been missed at
LEP.  The latter can greatly alter the experimental search strategy
for the Higgs and supersymmetry (SUSY) at the LHC.

In this paper, we will pursue the suggestion of \cite{ckr} and
\cite{ckr2}, that the Higgs can decay into light gauginos, which in
turn decay via renormalizable lepton number violating couplings,
into jets plus neutrinos. This decay would have been missed at LEP
if the Higgs is between $85 - 100$ GeV, and the gauginos are less
than half the Higgs mass.  We will discuss the FermiLab constraints
on this scenario in this paper, as well as constraints on like sign
dilepton decays of the Higgs, which inevitably accompany the decays
with neutrinos.  We find that there are plausible models in which
the branching ratios for like sign dileptons are small enough to
evade the strong, model independent, bounds from FNAL.

Our purpose is to go beyond the work of \cite{ckr2} in two ways.
First of all, we incorporate the L violating mechanism for hiding
the Higgs into gauge mediated SUSY breaking models.  Secondly, we
also exploit the lepton number violating operators to generate the
neutrino masses.  The seesaw mechanism for generating neutrino
masses, requires one to introduce a new mass scale, an order of
magnitude or so below the unification scale $M_U \sim 2 \times
10^{16}$ GeV\footnote{It has been suggested that this scale arises
naturally, as $M_{seesaw} = \frac{M_U^2}{m_P}$ \cite{ds}.}.
Renormalizable lepton number violating operators in SUSY can provide
a natural alternative \cite{numassinRPVSUSY}.   Our aim is to see
whether this can be combined with gauge mediation and simultaneously
hide the Higgs.

We will find that certain restrictions must be placed on the L
violating operators in order to achieve all of these goals.  Most of
our considerations are quite general, but we will specialize to the
Pentagon model \cite{pentagon} in order to investigate whether an
appropriate discrete symmetry can be found, which automatically
implies these restrictions.

Our attitude toward the magnitude of the possible L violating
operators is influenced by our knowledge of the Yukawa couplings in
the standard model.   Many of these are surprisingly small.  Given
the strong constraints on flavor changing neutral currents, we think
that the most plausible explanation of Yukawa textures is the
Froggatt-Nielsen mechanism \cite{FN} operating near the unification
scale.  It then seems clear that the flavor structure of L violating
operators will be similarly constrained.  Rather than trying to
formulate a full high energy theory of these textures, we merely
take away the lesson that dimensionless L violating couplings might
be anomalously small, and that one of them might be much larger than
all the others.

The MSSM also contains dimension two L violating operators,
analogous to the $\mu$ term, with $H_d$ replaced with a linear
combination of $L_i$.   Clearly, an explanation of the magnitude of
the dimension two parameter is necessary to a complete low energy
theory. We will adopt the philosophy of the NMSSM, in which this
parameter is the vacuum expectation value (VEV) of a low energy
singlet, and the bare dimension two couplings are forbidden by a
discrete symmetry.

\section{Constraints on $h \rightarrow \tilde{X} \tilde{X} :
\tilde{X} \rightarrow \nu + q + \bar{d} $}

We want to investigate the LEP bound on the Higgs mass in the MSSM
where the lightest Higgs boson is produced by Higgssthralung of the
Z boson (or maybe through Z or W-fusion processes).  The cascade
decay we are interested in consists of the decay of the lightest
Higgs to two next-to-lightest SUSY particle (NLSP) neutralinos
followed by an R-parity violating (RPV) decay of each neutralino to
one lepton plus two quarks.  We do the computation in the
narrow-width limit where the cascade is divided into a two-body
decay and two three-body decays.

\subsection{BR of the lightest Higgs to two neutralinos}

The partial decay width $\Gamma(h^0\rightarrow\chi_i^0\chi_j^0)$ is
given by
\begin{equation}
\Gamma(h^0\rightarrow\chi_i^0\chi_j^0)=\frac{\lambda^{1/2}(m_{h^0}^2,m_{\chi_i^0}^2,m_{\chi_j^0}^2)}{16\pi
m_{h^0}^3\times2^{\delta_{ij}}}\left(2|Y^{ij}|^2(m_{h^0}^2-m_{\chi_i^0}^2-m_{\chi_j^0}^2)-2[(Y^{ij})^2+(Y^{ij*})^2]m_{\chi_i^0}m_{\chi_j^0}\right)
\end{equation}
where
\begin{eqnarray}
\lambda(x,y,z) &=& x^2+y^2+z^2-2xy-2xz-2yz\\
Y^{ij} &=&
\frac{1}{2}(-N_{i3}^*\sin\alpha-N_{i4}^*\cos\alpha)(gN_{j2}^*-g'N_{j1}^*)+\{i\leftrightarrow
j\}.
\end{eqnarray}
Here $N$ diagonalizes the neutralino mass matrix $M_{\chi^0}$ which
can be written at tree level as
\begin{equation}
M_{\chi^0}=\left(
\begin{array}{cccc}
M_1 & 0 & -g'v_d/\sqrt{2} & g'v_u/\sqrt{2}\\
0 & M_2 & gv_d/\sqrt{2} & -gv_u/\sqrt{2}\\
-g'v_d/\sqrt{2} & gv_d/\sqrt{2} & 0 & -\mu\\
g'v_u/\sqrt{2} & -gv_u/\sqrt{2} & -\mu & 0
\end{array}
\right)
\end{equation}
where
$N^*M_{\chi^0}N^{-1}=\mbox{diag}(m_{\chi_1^0},m_{\chi_2^0},m_{\chi_3^0},m_{\chi_4^0})$
with $|m_{\chi_1^0}|<|m_{\chi_2^0}|<|m_{\chi_3^0}|<|m_{\chi_4^0}|$.

The total decay width is expected to be dominated by decays of the
lightest Higgs to neutralinos (when kinematically allowed), thus the
branching ratio can be approximated as
\begin{equation}
\mbox{BR}(h^0\rightarrow\chi_i^0\chi_j^0)=\frac{\Gamma(h^0\rightarrow\chi_i^0\chi_j^0)}{\Gamma(h^0\rightarrow\mbox{all})}\approx\frac{\Gamma(h^0\rightarrow\chi_i^0\chi_j^0)}{\Gamma(h^0\rightarrow\mbox{SM})+\Gamma(h^0\rightarrow\mbox{neutralinos})}.
\end{equation}

\subsection{BR of the neutralino to one lepton plus two quarks}

The decay of the neutralino to one lepton plus two quarks occurs
through the R-parity violating vertex
$\lambda_{ijk}'\epsilon_{ab}L_i^aQ_j^b\bar{D}_k\subset\mathcal{W}$.
Since squarks are assumed much heavier than sleptons, decays with
off-shell squarks are sub-dominant contributions to the partial
decay widths.  Moreover, assuming there is no mixing in the sfermion
sector $\tilde{f}_L$ and $\tilde{f}_R$ are mass eigenstates.  This
reduces the number of Feynman diagrams since only left-handed
sleptons and sneutrinos are relevant to the R-parity violating
vertex $\lambda_{ijk}'\epsilon_{ab}L_i^aQ_j^b\bar{D}_k$.  Thus only
decays with off-shell $\tilde{\ell}_{Li}$ or $\tilde{\nu}_i$ are
possible.  Finally, since the kinematically allowed final state standard 
model fermions (for a NSLP neutralino with mass $m_{\chi_1^0}\approx30$
GeV only the top quark is excluded as a final state fermion) are
much lighter than any sparticles one can compute the partial decay
widths in the limit of vanishing fermion masses.  This introduces a
maximal error of the order
$\mathcal{O}\big(\frac{m_b}{m_{\chi_1^0}}\big)\approx0.13$ for a
NLSP neutralino with mass $m_{\chi_1^0}\approx30$ GeV.  With these
assumptions, the partial decay width computations simplify greatly
and one can get analytical results.

Thus, with these assumptions, the partial decay widths
$\Gamma(\chi_l^0\rightarrow\ell_iu_j\bar{d}_k)$ and
$\Gamma(\chi_l^0\rightarrow\nu_id_j\bar{d}_k)$ are
\begin{eqnarray}
\Gamma(\chi_l^0\rightarrow\ell_iu_j\bar{d}_k)=\Gamma(\chi_l^0\rightarrow\bar{\ell}_i\bar{u}_jd_k) &=& \frac{N_cm_{\chi_l^0}(|c_1|^2+|c_2|^2)}{1024\pi^3}\left[6\rho-5+2(\rho-1)(3\rho-1)\ln\left(\frac{\rho-1}{\rho}\right)\right]\\
\Gamma(\chi_l^0\rightarrow\nu_id_j\bar{d}_k)=\Gamma(\chi_l^0\rightarrow\bar{\nu}_i\bar{d}_jd_k)
&=&
\frac{N_cm_{\chi_l^0}|c_1|^2}{1024\pi^3}\left[6\rho-5+2(\rho-1)(3\rho-1)\ln\left(\frac{\rho-1}{\rho}\right)\right]
\end{eqnarray}
where
\begin{eqnarray}
c_1 &=& \sqrt{2}\lambda_{ijk}'(gT_3^{f_i}N_{l2}^*+g'Y_{f_i}^HN_{l1}^*)\\
c_2 &=& \lambda_{ijk}'\frac{m_{\ell_i}}{v_d}N_{l3}\\
\rho &=& \left(\frac{m_{\tilde{f}_i}}{m_{\chi_l^0}}\right)^2.
\end{eqnarray}
Here $N_c=3$ is the number of colors and Martin's notation is used
for the hypercharge, i.e. $Q=T_3+Y^H$ \cite{m}.  Moreover, the first term in
$c_1$ represents the fermion/sfermion coupling to the wino, the
second term in $c_1$ represents the fermion/sfermion coupling to the
bino and $c_2$ represents the fermion/sfermion coupling to the
higgsino.  The following table reviews the needed hypercharges
\begin{equation}
\begin{array}{cccc}
\mbox{particle} & Y^H & T_3 & Q\\
\ell & -\frac{1}{2} & -\frac{1}{2} & -1\\
\nu & -\frac{1}{2} & \frac{1}{2} & 0\\
\end{array}
\end{equation}

Since the dominant decay of the lightest Higgs is expected to be
$h^0\rightarrow\chi_1^0\chi_1^0$, the NLSP neutralino decay  is the
one of interest.  The total decay width for the NLSP neutralino is
dominated by the kinematically allowed R-parity violating vertex
processes discussed above (the decay to the gravitino is
sub-dominant) thus
\begin{eqnarray}
\Gamma(\chi_1^0\rightarrow\mbox{all}) &\approx& \sum_{i,j,k}\left[\Gamma(\chi_1^0\rightarrow\ell_iu_j\bar{d}_k)+\Gamma(\chi_1^0\rightarrow\bar{\ell}_i\bar{u}_jd_k)+\Gamma(\chi_1^0\rightarrow\nu_id_j\bar{d}_k)+\Gamma(\chi_1^0\rightarrow\bar{\nu}_i\bar{d}_jd_k)\right]\\
 &=& 2\sum_{i,j,k}\left[\Gamma(\chi_1^0\rightarrow\ell_iu_j\bar{d}_k)+\Gamma(\chi_1^0\rightarrow\nu_id_j\bar{d}_k)\right]
\end{eqnarray}
where the sums over $u_j$ is limited to $j=\{1,2\}$ since the top
quark is not kinematically allowed.  Thus the relevant branching
ratios are
\begin{eqnarray}
\mbox{BR}(\chi_1^0\rightarrow\ell_iu_j\bar{d}_k) &=& \frac{\Gamma(\chi_l^0\rightarrow\ell_iu_j\bar{d}_k)}{\Gamma(\chi_1^0\rightarrow\mbox{all})}\\
\mbox{BR}(\chi_1^0\rightarrow\nu_id_j\bar{d}_k) &=& \frac{\Gamma(\chi_l^0\rightarrow\nu_id_j\bar{d}_k)}{\Gamma(\chi_1^0\rightarrow\mbox{all})}\\
\end{eqnarray}
with the same results for final state antiparticles.

\subsection{Total branching ratio in the regime $\tan \beta \sim 1$: Numerical example}\label{numerical}

The total branching ratio for one particular final state in the
cascade decay of interest is simply the product of the appropriate
branching ratios (in the narrow-width limit).  For example, for the
cascade decay
$h^0\rightarrow\chi_1^0\chi_1^0\rightarrow\ell_{i_1}u_{j_1}\bar{d}_{k_1}\ell_{i_2}u_{j_2}\bar{d}_{k_2}$
the total branching ratio is
\begin{equation}
\mbox{BR}(h^0\rightarrow\ell_{i_1}u_{j_1}\bar{d}_{k_1}\ell_{i_2}u_{j_2}\bar{d}_{k_2})=\mbox{BR}(h^0\rightarrow\chi_1^0\chi_1^0)\mbox{BR}(\chi_1^0\rightarrow\ell_{i_1}u_{j_1}\bar{d}_{k_1})\mbox{BR}(\chi_1^0\rightarrow\ell_{i_2}u_{j_2}\bar{d}_{k_2}).
\end{equation}
Our interest lies in decays with final state tau leptons and tau
neutrinos since these processes have not been studied extensively at
the LEP.

We will evaluate these branching ratios in the limit $\tan \beta
\sim 1$, as is predicted in the Pentagon model.  We do this here
because it turns out that in this regime, the like sign dilepton
contribution to Higgs decay is naturally suppressed.  The FNAL
bounds on this process will be more difficult to satisfy for larger
values of $\tan \beta$.

\begin{figure}[ht]\begin{center}\includegraphics[scale=0.65,angle=0]{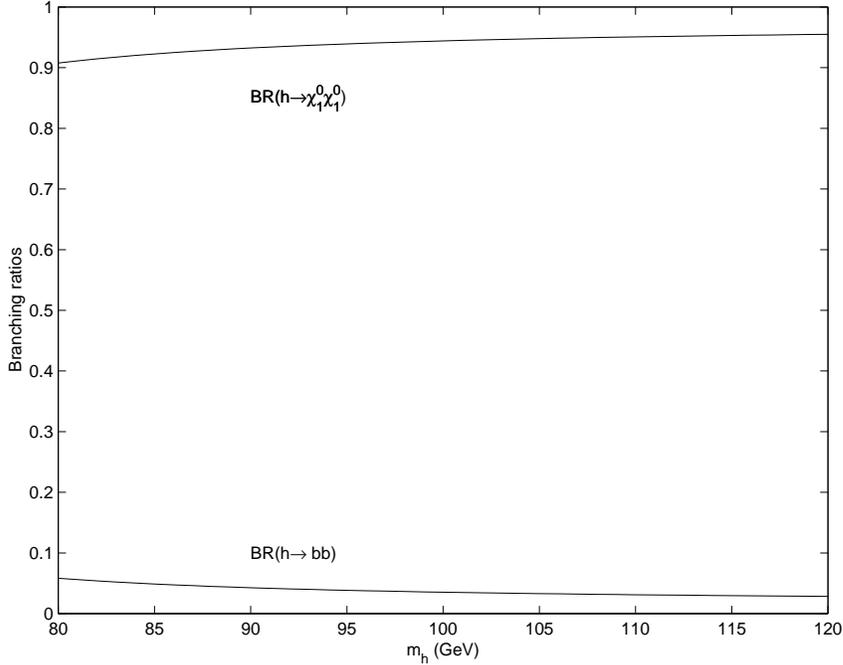}
\caption{\small{Branching ratio of $h^0\rightarrow\chi_1^0\chi_1^0$
(upper line) and $h^0\rightarrow b\bar{b}$ (lower line) as a
function of the Higgs mass (in GeV) for $M_1=50$ GeV, $M_2=250$ GeV, $\mu=+150$ GeV,
$\tan\beta=1$ and $\alpha=-\frac{\pi}{8}$.}}\label{Fig:BRHiggsNeutralinos}
\end{center}\end{figure}

Following \cite{ckr} we want the lightest neutralino $\chi_1^0$ to be
mostly bino.  This can be achieved with $M_1=50$ GeV, $M_2=250$ GeV and
$\mu=+150$ GeV which lead to the following masses
\begin{equation}
\begin{array}{c|cccc|cc}
\mbox{particle} & \chi_1^0 & \chi_2^0 & \chi_3^0 & \chi_4^0 & \chi_1^\pm & \chi_2^\pm\\
\hline
\mbox{mass (in GeV)} & 30 & 125 & 150 & 300 & 105 & 295
\end{array}
\end{equation}
and mixing matrix
\begin{equation}
N=\left(
\begin{array}{cccc}
-0.89 & 0.15 & -0.30 & 0.30\\
-0.44 & -0.48 & 0.54 & -0.54\\
0 & 0 & 0.71 & 0.71\\
-0.09 & 0.86 & 0.35 & -0.35\\
\end{array}
\right)
\end{equation}
Close to the Higgs decoupling limit this leads to
$\mbox{BR}(h^0\rightarrow\chi_1^0\chi_1^0)\approx0.8 - 0.9$ for $m_{h^0}$ between
$85 - 100$ GeV.  Figure \ref{Fig:BRHiggsNeutralinos} shows the branching ratios $\mbox{BR}(h^0\rightarrow\chi_1^0\chi_1^0)$
and $\mbox{BR}(h^0\rightarrow b\bar{b})$ as a function of the Higgs mass for a mixing angle
$\alpha=-\frac{\pi}{8}$.  Here, since $\tan\beta=1$, we go slightly away from the Higgs decoupling
limit in order to satisfy the experimental bound on $\xi^2\mbox{BR}(h^0\rightarrow b\bar{b})$
for $m_h$ as low as $85$ GeV \cite{LEPbound}.

Using PDG bounds \cite{PDG} on the masses of the sleptons, appropriate values
for the masses of the sneutrinos\footnote{For $\tan\beta=1$, theory
forces the sneutrinos to be almost degenerate with the sleptons.
However from the non-SM invisible width of the $Z$-boson sneutrinos
could be as light as $45$ GeV.}
\begin{equation}
\begin{array}{c|ccc|ccc}
\mbox{particle} & \tilde{\ell}_1=\tilde{e} & \tilde{\ell}_2=\tilde{\mu} & \tilde{\ell}_3=\tilde{\tau} & \tilde{\nu}_1=\tilde{\nu}_e & \tilde{\nu}_2=\tilde{\nu}_\mu & \tilde{\nu}_3=\tilde{\nu}_\tau\\
m\;\mbox{(in GeV)} & 73 & 94 & 81.9 & 75 & 75 & 75
\end{array}
\end{equation}
and for the R-parity violating coupling \cite{numassinRPVSUSY}
\begin{eqnarray}
\lambda_{ijk}' &=& 0\;\forall\;i=\{1,2\},\;j,k=\{1,2,3\}\\
(\lambda_{3jk}') &=& \left(
\begin{array}{ccc}
0.001 & 0.001 & 0\\
0.001 & 0.001 & 0\\
0 & 0 & 0
\end{array}
\right)
\end{eqnarray}
one obtains $\mbox{BR}(\chi_1^0\rightarrow\tau u_j\bar{d}_k)=0.019$
$\forall\;j,k=\{1,2\}$, $\mbox{BR}(\chi_1^0\rightarrow\nu_\tau
d_j\bar{d}_k)=0.11$ $\forall\;j,k=\{1,2\}$ and zero otherwise.
The difference between the branching ratios comes from the mixing
matrix $N$ and the quantum numbers $T_3$ and $Y^H$.  Indeed, since
the sneutrino is almost degenerate with the stau one can
forget about the $\rho$-dependent part of the decay width and focus
on the $c_1$-bino and wino contributions to the decay width (the
$c_2$-higgsino contributions to the decay width are small due to the
lepton mass suppression factor).  The $c_1$-bino contributions to
the branching ratio for $\tau$ and $\nu_\tau$ are the same
\emph{but} the $c_1$-wino contributions to the branching ratio for $\tau$
and $\nu_\tau$ have opposite sign.  From the mixing matrix one can
see that the contributions partly cancel for $\tau$ and add up for
$\nu_\tau$.  Though smaller, the $c_1$-wino contributions are about
a third of the $c_1$-bino contributions which leads to a suppression
\begin{equation}
\frac{\mbox{BR}(\chi_1^0\rightarrow\tau
u_j\bar{d}_k)}{\mbox{BR}(\chi_1^0\rightarrow\nu_\tau
d_j\bar{d}_k)}\approx\frac{|c_{1,\tau}|^2}{|c_{1,\nu_\tau}|^2}\approx\frac{(-1+3)^2}{(1+3)^2}\approx\frac{1}{4}.
\end{equation}
As shown before the branching ratios for antiparticles in the final
state are the same.

Thus, the total branching ratios for the following cascade decays
are
\begin{eqnarray}
\mbox{BR}(h^0\rightarrow\tau\tau+4\;\mbox{jets})\approx0.0055\\
\mbox{BR}(h^0\rightarrow\tau\nu_\tau+4\;\mbox{jets})\approx0.03\\
\mbox{BR}(h^0\rightarrow\nu_\tau\nu_\tau+4\;\mbox{jets})\approx0.17
\end{eqnarray}
with $m_{h^0}\approx90$ GeV and $\alpha=-\frac{\pi}{8}$.
Here any group of particles associated to
the neutralino decay products can be changed by its antiparticle
counterpart without changing the branching ratios.  Thus one obtains
same sign di-tau (or di-antitau) events with the same branching
ratio then tau-antitau events.

The total decay width of the Higgs and the NLSP neutralino are
\begin{eqnarray}
\Gamma(h^0\rightarrow\mbox{all}) &\approx& 0.02\;\mbox{GeV}\\
\Gamma(\chi_1^0\rightarrow\mbox{all}) &\approx& 0.11\cdot10^{-10}\;\mbox{GeV}
\end{eqnarray}
and this may lead to displaced vertices since $c\tau_{\chi_1^0\rightarrow\mbox{all}}\approx19\;\mbox{$\mu$m}$.

\subsection{Gaugino mass relationships and the chargino mass bound}

There is a strict lower mass bound of 102.7 GeV set on charginos
that decay through RPV operators.  We note that if we assume the
minimal gauge mediated prediction for the gaugino mass ratio, a
chargino of 102.7 GeV would make it impossible to have neutralinos
less than 50 GeV.  In order to allow neutralinos of less than half
the Higgs mass and satisfy the chargino mass bound, we must alter
the minimal gauge mediation predictions for gaugino masses. We
present here a simple method proposed by \cite{cdfm} of adding to the
hidden sector multiple scalar fields which get supersymmetric and
non-supersymmetric masses, and coupling them with different
strengths to parts of a single $10$ and $\overline{10}$ of messengers.
The superpotential is thus

\begin{equation}
W= r_i X_i u \overline{u}+ \gamma_i X_i q \overline{q} + \lambda_i X_i l\overline{l}
\end{equation}
for
\begin{equation}
X_i = x_i + \theta^2 F_i
\end{equation}

The resulting  gaugino eigenstates are determined by three mass
parameters instead of a single mass parameter in the minimal case.
The parameters are,

\begin{equation}
\Lambda_l = \frac{\lambda_i F_i }{\lambda_i x_i },\;\;\; \Lambda_q =  \frac{\gamma_i F_i}{\gamma_i x_i },\;\;\; \Lambda_u = \frac{r_i F_i }{r_i x_i}
\end{equation}
and the resulting gaugino mass parameters are
\begin{equation}
M_1 = \frac{1}{2} \frac{\alpha_1}{4 \pi}\left(\frac{4}{3} \Lambda_q + 2\Lambda_l+ \frac{8}{3} \Lambda_u\right),\;\;\; M_2 =\frac{3}{2} \frac{\alpha_2}{4 \pi}\Lambda_q,\;\;\; M_3= \frac{1}{2}\frac{\alpha_3}{4 \pi}(\Lambda_u +2\Lambda_q)
\end{equation}
Here we have changed the ordinary gaugino mass ratio
with minimal extra structure and it is easy to get a $M_1$ much lighter than $M_2$.

\section{Tevatron bounds on jets + $\displaystyle{\not E}_T$}

The Tevatron has searched \cite{jetsMET} for jets plus missing
energy coming from top-anti-top production, including four jet
events. We must inquire whether the gaugino decays we have described
should have been seen in these searches.  This particular search
places limits on the $t\overline{t}$ production cross section by
looking for multi jet events with 2 b quarks in the final state and
with missing energy, which is not found in the b quark direction. In
principle this looks very similar to our signal and we must ask if
we are constrained by this search.

There are several reasons to expect this search to be insensitive to
our decay.  The most important remark is that the Higgs production
cross section in the regime of interest is a few picobarns
\cite{rainwater}.  The same can be said for the direct neutralino
production cross section.   The two sigma error bars for the
$t\overline{t}$ production cross section at Tevatron are about 3 pb.
Thus, the jet plus missing energy events found by Tevatron, which
are all consistent with coming from top, probably put very weak
bounds on $LQ\bar{D}$ couplings.  In addition, since the Higgs is
produced close to threshold at Tevatron, the decay products are all
relatively soft.  The most missing energy we expect is from final
state with two neutrinos of about 15 GeV.  Since the decay products
are isotropically distributed, it is unlikely that the two neutrinos
will be found in the same hemisphere, and the total missing energy
vector will not be large enough to pass missing energy cuts in this
search.  Finally there is a question about how many b quarks are in
the final state.  The Fermilab search insisted on two b tagged jets
as one of their primary cuts.   Since the flavor structure of the $L
Q \bar{D}$ couplings is undetermined, we could easily construct
models, which further suppressed b jets.  Thus gaugino decays
through lepton number violation could have easily escaped detection.
However,  since the missing energy threshold is small, if there are
too many tagged b's in the event it will be discarded as QCD
background. Therefore events with four b's in the final state are
less likely to be picked up by this search.

That being said, the kinematics of the gaugino decays are
sufficiently different from the top decays, that one could imagine
finding them in a dedicated search. One could attempt a search
without b tagging, but simple searches for jets plus significant
missing energy are very difficult.  The best chance might be to
assume that the Higgs decays have the maximal number of b's in the
final state, and to modify the Tevatron 4b Higgs search to be
sensitive to missing energy \cite{Jason}.  In the standard 4b search
the Higgs, which decays to $b\overline{b}$, is produced in
association with one or two b quarks \cite{CDF4b}.  This search
required at least 3 b-tags, but overall it was not very sensitive
since it was sensitive to the 3b background.  One may imagine a
similar search which requires multiple b tags and a missing energy
cut. Problems remain: the missing energy must be more than what we
would expect from semi-leptonic b decay, and more than what would
result from the mis-measurement of a 4 b event with no missing
energy. Passing these cuts with low energy neutrinos may be a
problem. Searches at LHC seem to be even more problematic, but we
note that Kaplan and Rehermann have proposed searching for Higgs
decays through neutrino LSPs into multi jet final states using the
LHCb experiment \cite{LHCb}.  LHCb catches events highly boosted in
the forward direction, has maximal b acceptance, and has a $p_T$
trigger which can be as low as 2 GeV. As in the B violating 6 jet
decays of the Higgs \cite{ckr}, it may be possible to search for
lepton number violating decays of the gaugino at LHCb, with missing
energy and multiple b tags.

\section{Neutrino masses}

The lepton number violating operators, which we have invoked to hide
the decays of a light Higgs boson, might also be the source of
neutrino masses.  There is a large literature (see
\cite{neutrinomasses} and references therein) on the use of
renormalizable L violating terms in the MSSM to generate neutrino
masses.  Indeed, some of the strongest constraints on the L
violating couplings we have used come from the requirement that the
neutrino masses they generate not be too large.  For instance, the
numerical example of section \ref{numerical} leads to a
$\lambda'\lambda'$ loop neutrino mass of about $1.5\cdot10^{-3}$ eV 
for squark of about $250$ GeV \cite{neutrinomasses}.  A survey of the literature indicates that
bilinear L violating terms of the form $L_i H_u$ are the dominant
source of neutrino masses in a generic model. However, to be
consistent one should require that all B and L violating terms which
could lead to unobserved processes are forbidden by a symmetry.  We
do not know how to make a general analysis of such symmetries
without committing ourselves to a specific model.  Thus we will
restrict our attention to the Pentagon model \cite{pentagon}, though
we expect that a similar analysis could be done for any specific
model of gauge mediation.  We will find, that within the context of
the Pentagon model, the symmetries we utilize will forbid terms of
the form $L_i H_u$ but allow $L_i H_u S$ (where $S$ is the singlet
of the Pentagon).  If $S$ has a vacuum expectation value (VEV), this
will generate a tree level mass for one neutrino.  The dominant
contribution to the other two neutrino masses comes from loop
corrections involving the $LQ\bar{D}$ couplings that hide the Higgs
decay. There is thus a potential understanding of a $2-1$ hierarchy
among the three neutrino masses, as seems to be indicated by
experiment. However, we emphasize that both the magnitude of the
$L_i H_u S$ term, and the $LQ \bar{D}$ couplings is determined by
high energy physics beyond the range of the present analysis.
Therefore, a proper understanding of the structure of the neutrino
mass matrix really requires unification scale physics.

The original Pentagon model was designed to eliminate all baryon and
lepton number violating operators of dimension $\leq 5$, except for
the neutrino seesaw term.  This led to a $\mathbbm{Z}_4$ R symmetry with two
possible generation independent charge assignments. In order to
admit renormalizable lepton number violating terms we must change
the symmetry and the charge assignments.  We will assume an R
symmetry group $\mathbbm{Z}_N$. Therefore in the following all
equations are understood modulo $N$ and the R-charge of a given
field is denoted by the name of the field itself.

\subsection{Independent R-charges}

The aim of this subsection is to express the R-charges of all the
fields of the Pentagon in terms of the R-charges of a minimal set of fields. The
appropriate restricted set is somewhat arbitrary but a rather
convenient one comes naturally from the model.  First, the crucial
$SP\tilde{P}$ and $SH_uH_d$ terms lead to
\begin{eqnarray}
SP\tilde{P} &\Rightarrow& P+\tilde{P}=2-S\\
SH_uH_d &\Rightarrow& H_u=2-S-H_d.
\end{eqnarray}
The important Yukawa couplings give
\begin{eqnarray}
LH_d\bar{E} &\Rightarrow& \bar{E}=2-L-H_d\\
QH_u\bar{U} &\Rightarrow& \bar{U}=2-Q-H_u\\
QH_d\bar{D} &\Rightarrow& \bar{D}=2-Q-H_d.
\end{eqnarray}
Thus one can rewrite everything as a function of the (extended)
restricted set $\{S,L,Q,H_d\}$ as
\begin{eqnarray}
P+\tilde{P} &=& 2-S\\
H_u &=& 2-S-H_d\\
\bar{E} &=& 2-L-H_d\\
\bar{U} &=& S-Q+H_d\\
\bar{D} &=& 2-Q-H_d.
\end{eqnarray}
This set is dubbed extended since anomaly conditions will generate
relations between the four different R-charges of the set.

\subsubsection{Anomaly conditions}

The anomaly conditions of the Pentagon model are
\begin{eqnarray}
SU(5)_P &\Rightarrow& 5(P+\tilde{P})=0\\
SU(3)_C &\Rightarrow& 6Q+3(\bar{U}+\bar{D})+5(P+\tilde{P})=0\\
SU(2)_L &\Rightarrow& H_u+H_d+9Q+3L+5(P+\tilde{P})=0.
\end{eqnarray}
Using the relations obtained from the restricted set of independent
R-charges $\{S,L,Q,H_d\}$ these can be rewritten as
\begin{eqnarray}
SU(5)_P &\Rightarrow& 5(S-2)=0\\
SU(3)_C &\Rightarrow& 3(S+2)=0\\
SU(2)_L &\Rightarrow& 2-S+9Q+3L=0.
\end{eqnarray}
The last anomaly condition leads to an unextended restricted set of
independent R-charges by removing one R-charge in the extended
restricted set.  Due to the modulo $N$ form of the equations the
easiest one to remove is $S$ but it is more convenient to keep
everything written in function of the extended restricted set
$\{S,L,Q,H_d\}$.  Indeed one can easily solve the anomaly conditions
in function of $S$.  Thus it is more practical to eliminate $Q$
instead as shown later.  The first two anomaly conditions can be
combined as
\begin{equation}
0=5(S-2)=3(S+2)+2(S-8)=2(S-8).
\end{equation}
For $N=2n+1$ one has $S=8$ and the first two anomaly conditions
force $N|30$ thus $N=\{3,5,15\}$.  For $N=2n$ one has $S=8$ or
$S=8-n$.  For the case $S=8$ the first two anomaly conditions force
$N|30$ thus $N=\{2,6,10,30\}$.  For the case $S=8-n$ the first two
anomaly conditions lead to
\begin{eqnarray}
N &|& (30-5n)\\
N &|& (30-3n)
\end{eqnarray}
thus $N=\{4,12,20,60\}$.

\subsection{Superpotential terms and RPV terms}

Using the extended restricted set the possible $S^3$ superpotential
term gives
\begin{equation}
S^3\Rightarrow3S-2
\end{equation}
where the RHS has to be zero modulo $N$ if and only if the term is
allowed.  For the RPV terms, the trilinear lepton number violating
(TLNV) terms (including the useful $SLH_u$) give
\begin{eqnarray}
LL\bar{E} &\Rightarrow& 2L+(2-L-H_d)-2=L-H_d\\
LQ\bar{D} &\Rightarrow& L+Q+(2-Q-H_d)-2=L-H_d\\
SLH_u &\Rightarrow& S+L+(2-S-H_d)-2=L-H_d.
\end{eqnarray}
and the bilinear lepton number violating (BLNV) term $LH_u$ leads to
\begin{equation}
LH_u\Rightarrow L+(2-S-H_d)-2=L-H_d-S.
\end{equation}
Finally the trilinear baryon number violating (TBNV) term gives
\begin{equation}
\bar{U}\bar{D}\bar{D}\Rightarrow(S-Q+H_d)+2(2-Q-H_d)-2 = S-3Q-H_d+2.
\end{equation}
The no-go theorem here states that one cannot allow only specific
TLNV terms since all TLNV terms are allowed when anyone is allowed.

\subsection{Dimension five baryon number violating operators}

Dimension five baryon number violating (D5BNV) operators and D-terms
lead to
\begin{eqnarray}
QQQL &\Rightarrow& 3Q+L-2\\
QQQH_d &\Rightarrow& 3Q+H_d-2\\
\bar{U}\bar{U}\bar{D}\bar{E} &\Rightarrow& 2(S-Q+H_d)+(2-Q-H_d)+(2-L-H_d)-2=2S-3Q-L+2\\
\mbox{D-term} &\Rightarrow& Q+\bar{U}-L=Q+(S-Q+H_d)-L=S-L+H_d\\
\mbox{D-term} &\Rightarrow&
\bar{U}+\bar{E}-\bar{D}=(S-Q+H_d)+(2-L-H_d)-(2-Q-H_d)=S-L+H_d
\end{eqnarray}
where the last two equations come from D-terms.

\subsection{Overall solutions}

From the last two sections one can group together terms that lead to
the same equation in function of the R-charges of the extended
restricted set.  One has seven different sets labeled $G_1$ to
$G_7$,
\begin{eqnarray}
G_1=\{S^3\} &\Rightarrow& 3S-2\\
G_2=\{LL\bar{E},LQ\bar{D},SLH_u\} &\Rightarrow& L-H_d\\
G_3=\{LH_u,\mbox{D-terms}\} &\Rightarrow& L-H_d-S\\
G_4=\{\bar{U}\bar{D}\bar{D}\} &\Rightarrow& S-3Q-H_d+2\\
G_5=\{QQQL\} &\Rightarrow& 3Q+L-2\\
G_6=\{QQQH_d\} &\Rightarrow& 3Q+H_d-2\\
G_7=\{\bar{U}\bar{U}\bar{D}\bar{E}\} &\Rightarrow& 2S-3Q-L+2
\end{eqnarray}
where group $G_2$ consists of all TLNV terms exclusively, group $G_4$ of the
TBNV term and groups $G_5$ to $G_7$ of D5BNV terms.  Using these
sets and the extra $SU(2)_L$ anomaly relation $2-S+9Q+3L$ to
eliminate $Q$ the solutions are
\begin{equation}
\begin{array}{c|cccccccc}
N & S & SU(2)_L & G_1 & G_3 & G_4 & G_5 & G_6 & G_7\\
\hline
2 & 0 & Q=L & 0 & L-H_d & L-H_d & 0 & L-H_d & 0\\
3 & 2 & \mbox{none} & 1 & L-H_d-2 & 1-H_d & L-2 & H_d-2 & -L\\
4 & 2 & Q=L & 0 & L-H_d-2 & L-H_d & 2 & -(L-H_d-2) & 2\\
5 & 3 & Q=3L-1 & 2 & L-H_d-3 & L-H_d-2 & 0 & -(L-H_d) & 1\\
6 & 2 & 3Q=3L & 4 & L-H_d-2 & 3L-H_d-2 & -2(L+1) & 3L+H_d-2 & 2L\\
10 & 8 & Q=3L+4 & 2 & L-H_d-8 & L-H_d-2 & 0 & -(L-H_d) & 6\\
12 & 2 & 3Q=3L & 4 & L-H_d-2 & 4-3L-H_d & 4L-2 & 3L+H_d-2 & -4L+6\\
15 & 8 & 9Q=6-3L & 7 & L-H_d-8 & 10-3Q-H_d & 3Q+L-2 & 3Q+H_d-2 & -3Q-L+3\\
 & 8 & 3Q=2-L & 7 & L-H_d-8 & L-H_d+8 & 0 & -(L-H_d) & 1\\
 & 8 & 3Q=7-L & 7 & L-H_d-8 & L-H_d+3 & 5 & -(L-H_d-5) & -4\\
 & 8 & 3Q=12-L & 7 & L-H_d-8 & L-H_d-2 & 10 & -(L-H_d-10) & -9\\
20 & 18 & 9Q=16-3L & 12 & L-H_d-18 & -3Q-H_d & 3Q+L-2 & 3Q+H_d-2 & -3Q-L+18\\
 & 18 & 3Q=12-L & 12 & L-H_d-18 & L-H_d-12 & 10 & -(L-H_d-10) & 6\\
30 & 8 & 9Q=6-3L & 22 & L-H_d-8 & 10-3Q-H_d & 3Q+L-2 & 3Q+H_d-2 & -3Q-L+18\\
 & 8 & 3Q=2-L & 22 & L-H_d-8 & L-H_d+8 & 0 & -(L-H_d) & 16\\
 & 8 & 3Q=12-L & 22 & L-H_d-8 & L-H_d-2 & 10 & -(L-H_d-10) & 6\\
 & 8 & 3Q=22-L & 22 & L-H_d-8 & L-H_d-12 & 20 & -(L-H_d-20) & -4\\
60 & 38 & 9Q=36-3L & 52 & L-H_d-38 & 40-3Q-H_d & 3Q+L-2 & 3Q+H_d-2 & -3Q-L+18\\
 & 38 & 3Q=12-L & 52 & L-H_d-38 & L-H_d+28 & 10 & -(L-H_d-10) & 6\\
 & 38 & 3Q=32-L & 52 & L-H_d-38 & L-H_d+8 & 30 & -(L-H_d-30) & -14\\
 & 38 & 3Q=52-L & 52 & L-H_d-38 & L-H_d-12 & 50 & -(L-H_d-50) & -34
\end{array}
\end{equation}
where the TLNV set $G_2\Rightarrow L-H_d$ does not simplify.  Notice
that no extra relation comes from the $SU(2)_L$ anomaly condition
for $N=3$.  The removal of $Q$ from the extended restricted set is
more subtle for the cases $N=\{15,20,30,60\}$.  For $N=15$ one has
$9Q+3L=15k+6\Rightarrow3Q+L=5k+2$ ($k\in\mathbbm{Z}$) thus
$3Q=\{2-L,7-L,12-L\}$.  For $N=20$ one has
$9Q+3L=20k'+16\Rightarrow3Q+L=\frac{1}{3}(20k'+16)=20k+12$ with
$k'=3k+1$ ($k\in\mathbbm{Z}$) thus $3Q=12-L$.  For $N=30$ one has
$9Q+3L=30k+6\Rightarrow3Q+L=10k+2$ ($k\in\mathbbm{Z}$) thus
$3Q=\{2-L,12-L,22-L\}$.  Finally for $N=60$ one has
$9Q+3L=60k+36\Rightarrow3Q+L=20k+12$ ($k\in\mathbbm{Z}$) thus
$3Q=\{12-L,32-L,52-L\}$.

Looking at the previous table one sees that only the cases
$N=\{2,4\}$ allow for the $S^3$ term set $G_1$.  In the case of
interest to us, i.e. allowing TLNV terms set $G_2$ (thus $H_d=L$)
while prohibiting sets $G_3$ to $G_7$, one can find a solution only
for $N=\{12,15,20,30,60\}$ (notice that $G_3$ is not a problem
unless $N=2$).  For example the case $N=3$ is not a solution since
prohibiting $G_5$ and $G_7$ forces $L=1$ which allows the unwanted
$G_4$ while the case $N=12$ is a solution since the sets $G_5$,
$G_6$ and $G_7$ do not constrain $L$ but $G_4$ forces
$L\neq\{1,4,7,10\}$ which is possible.  The specific R-charges for
the five possible cases are then computable.  In this framework it
is therefore impossible to allow only TLNV terms set $G_2$ along
with the $S^3$ term set $G_1$.  It is however possible to allow only
TLNV terms set $G_2$.

\subsection{Constraints on $\langle S \rangle$}

In light of the previous computations one can engineer the
appropriate Pentagon superpotential
\begin{equation}
\mathcal{W}=(m_{\mbox{{\tiny ISS}}}+g_SSY)P\tilde{P}+g_\mu
SH_uH_d+\lambda_LH_dL\bar{E}+\lambda_uH_uQ\bar{U}+\lambda_dH_dQ\bar{D}+\frac{1}{2}\lambda
LL\bar{E}+\lambda'LQ\bar{D}+g_\epsilon SLH_u.
\end{equation}
If $S$ gets a VEV then \emph{one} neutrino mass is mostly due to the
$SLH_u$ term while the Higgs is hidden by the $LQ\bar{D}$ term. This
comes from the specific form of the tree level neutrino mass matrix
($\mbox{rank}=1$) and thus only one neutrino is massive which is
good to generate a hierarchy.  Loop diagrams from the $LL\bar{E}$
and $LQ\bar{D}$ terms give masses to the other neutrinos (see \cite{neutrinomasses}).
There is also an effective $\mu$ term and thus no light higgsinos.

On the other hand, it may be a challenge to give a VEV to $S$ if
there is no $S^3$ term.  If there is no $S$ VEV then we could get
both neutrino masses and Higgs decay to jets plus missing energy
from the $LQ\bar{D}$ term, but we are likely to have an unacceptable
light higgsino.  A model without a VEV for $S$ could generate all
neutrino masses through loops involving the $LQ\bar{D}$ couplings,
and could hide the Higgs via these same couplings.   However, it
would probably have an unacceptable light higgsino.

\section{Conclusions}

We have seen that gauge mediated models with lepton number violation
can in principle hide the Higgs and generate an acceptable neutrino
mass spectrum simultaneously.   Our attempt to find a model in which
the appropriate couplings followed from a discrete symmetry of the
low energy theory was not completely successful.

The problem we encountered was specific to embedding the lepton
violating scenario into the framework of the Pentagon model, but we
anticipate some general features.  In particular, it seems hard to
find models where low energy symmetries allow $LQ \bar{D}$
operators, but forbid $LL\bar{E}$ operators.  One has to rely on a
high energy Froggatt-Nielsen mechanism, combined with SUSY
non-renormalization theorems, to explain the suppression of the
latter, which are significantly more constrained.

There will also be an inevitable connection between the origin of
neutrino masses in R-parity violating models and the $\mu$ term of
the MSSM.   Our analysis indicates that it may be hard to explain
the value of $\mu$ in terms of a low energy singlet VEV in these
lepton number violating models.

Nonetheless, we think that gauge mediated models with renormalizable
lepton number violation could offer considerable insight into two
puzzles of the standard model.   We have barely scratched the
surface of this general class of models, and they deserve further
investigation.

\section{Acknowledgments}

We would like to thank Michael Dine, Lance Dixon, Eva Halikadakis,
Amit Lath, M. Maggi, Jason Nielsen, and Scott Thomas, for important
conversations.

This research was supported in part by DOE grant number
DE-FG03-92ER40689.

\end{document}